# Advanced Characterization-Informed Framework and Quantitative Insight to Irradiated Annular U-10Zr Metallic Fuels

Fei Xu[*]; Lu Cai[*]; Daniele Salvato; Fidelma Dilemma; Luca Capriotti; Tiankai Yao[#]

[*]These authors contributed equally

Idaho National Laboratory, Idaho Falls, ID, 83401, USA

[#]Corresponding authors

Tiankai Yao, Email: tiankai.yao@inl.gov

**Abstract**

U-10Zr-based metallic nuclear fuel is a promising fuel candidate for next-generation sodium-cooled fast reactors. Idaho National Laboratory's research experience for this type of fuel dates back to the 1960s. Idaho National Laboratory researchers have accumulated a considerable amount of experience and knowledge regarding fuel performance at the engineering scale. The limitation of advanced characterization and lack of proper data analysis tools prevented a mechanistic understanding of fuel microstructure evolution and properties degradation during irradiation. This paper proposed a new workflow, coupled with domain knowledge obtained by advanced post-irradiation examination methods, to provide unprecedented and quantified insights into the fission gas bubbles and pores, and lanthanide distribution in an annular fuel irradiated in the Advanced Test Reactor. In the study, researchers identify and confirm that the Zr-bearing secondary phases exist and generate the quantitative ratios of seven microstructures along the thermal gradient. Moreover, the distributions of fission gas bubbles on two samples of U-10Zr advanced fuels were quantitatively compared. Conclusive findings were obtained and allowed for evaluation of the lanthanide transportation through connected bubbles based on approximately 67,000 fission gas bubbles of the two advanced samples.

1. **Introduction**

Machine-learning (including deep learning), a type of artificial intelligence (AI) used to predict an outcome based on historical data, has become increasingly important in materials science. The use of machine-learning resulted in the prediction of material structure and/or properties, and created insights in the classic relationships of material structure – processing – property – performance for different material systems.[1,2] Machine-learning is one of the most promising tools to accelerate research, development, qualification, and commercial licensing of nuclear materials.[3] The research and eventually deployment for commercial use of nuclear materials, which need to function under irradiation at extreme environments (high temperature and/or high pressure and/or highly corrosive environments, etc.), are very expensive and time-costly, because number of out-of-file tests, in-reactor (irradiation) tests, and post-irradiation examination (PIE) are required to understand and be able to predict the material performance and degradation.[4] The goal of this present study is to use machine-learning to aid/bridge the mechanistic understanding of U-10wt.% Zr (U-10Zr) based metallic fuel, a primary candidate for next generation sodium cooled fast reactors (SFRs),[5] and eventually accelerate fuel qualification and licensing for commercial use.

U-10Zr fuels were used and tested extensively in test reactors, such as Experimental Breeder Reactor II (EBR-II) and Fast Flux Test Facility (FFTF) from the 1960s to the 1990s.[6,7] A lot of new PIE data within the sub-nanometer to micrometer scale has been recently generated.[8-10] Two main challenges of this fuel material are zirconium redistribution (Fuel Constituent Migration),[11] and fuel-cladding chemical interactions (FCCI).[12] FCCI is a chemical reaction between nuclear fuel and cladding (a thin-walled metal as first barrier for retention of fission products), which creates an important challenge to overcome in metallic fuels.[7,13] FCCI is generally dominated by the reactions between lanthanide fission products and iron-based cladding, resulting in eroded cladding with deteriorated mechanical properties, which adversely impacts the fuel integrity and performance. Lanthanides are transported from the hottest fuel center to the cold inner cladding through interconnected pores formed by fission gas bubbles.[9,14] Thus, the understanding of porosity distribution is essential for the slowdown or stop of the lanthanide transportation, and thereby mitigating FCCI. Fission gas bubbles, however, can lead to a 35% reduction of thermal conductivity for metallic fuel.[15] Fission gas bubbles show different sizes and distribution patterns along a radial thermal gradient inside irradiated U. The gas bubbles are primarily round and significantly smaller in the hot center zone.[16]

Another key factor is constituent redistribution along the fuel cross-section. Fuel constituent radial redistribution during irradiation changes the local fuel composition and results in fuel melting temperature.[8] The temperature gradient within the U-10Zr fuel, as well as irradiation enhanced diffusion, causes zirconium and uranium to migrate in the opposite directions, resulting in different zones with different characteristic microstructures and crystal structures. The zirconium redistribution can be beneficial, because it increases the solidus temperature in the hottest fuel region when Zr migrates to the fuel center.[6] However, the zirconium redistribution may also have adverse impacts on the fuel performance as it may influence/degrade the material properties (e.g., thermal conductivity) of the fuel.[7,10] It is essential to achieve a comprehensive understanding of the zirconium redistribution (phase distribution) in order to better predict fuel performance.[11] A U-10Zr fuel cross-section would host Zr redistribution in three major phases: $\gamma$ phase (a continuous body centered cubic solid solution between U and Zr), noted as the (U, Zr) matrix; $\alpha$-U matrix; and $\alpha$-U + $UZr_2$. The Zr redistributes across the pin radius and forms up to three distinctive zones by different U/Zr contents.[17]

In our previous study, machine-learning has been successfully applied to characterize and classify pores and gain insights into the lanthanide transportation of an advanced U-10Zr annular fuel (AF1).[17] In this paper, another advanced U-10Zr fuel, named AF2 was investigated. The pipeline of the proposed method is shown in Figure 1. Two major processes were involved in this study: 1) analyzing the distributions of two major elements U and Zr on six image patches and 2) exploring the fission gas bubble distribution along the thermal degradation from hot region to colder region close to the cladding region on a partial cross-section. In the process of fission gas bubble analysis, the $UZr_2$ phase is identified; then bubbles are segmented from the background (detailed in Sections 2.2 and 2.3); and then a trained Decision Tree classifier is applied to cluster the bubbles into categories such as large/intermediate/small size and connected/isolated (as described in Section 2.4). Lastly, the quantitative results of the new advanced fuel are generated and used to obtain conclusive findings by comparing the two advanced fuels.

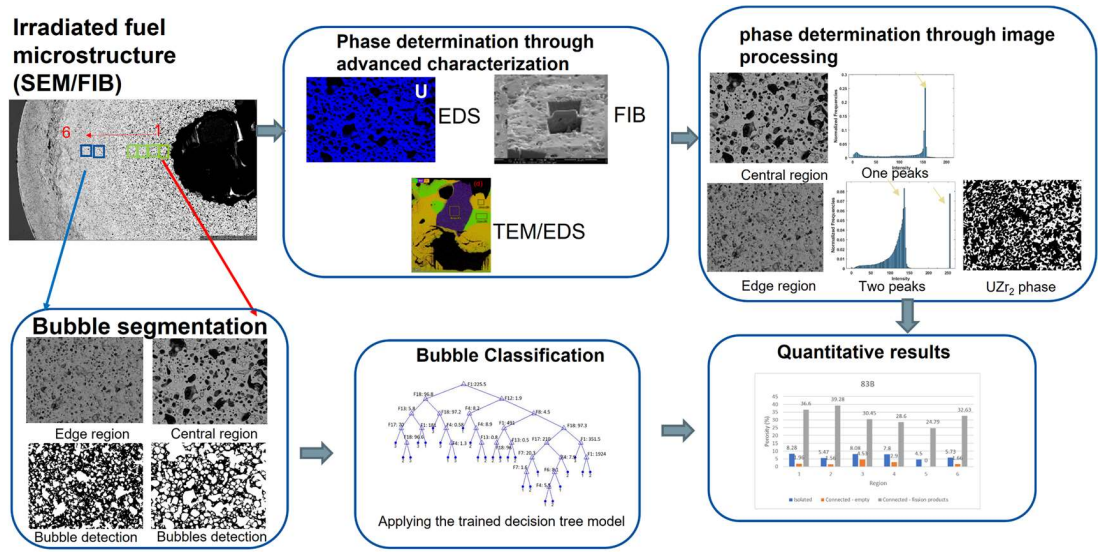

Figure 1. Workflow of the proposed method.

## 2. The Proposed Method

### 2.1. Experimental Data

Idaho National Laboratory (INL) has been the leading national laboratory for research and development (R&D) of metallic fuel, including multiple fuel forms and post-irradiation characterization. Advanced characterization techniques, e.g., focused ion beam (FIB) sampling, transmission electron microscopy (TEM) characterization, and local thermal conductivity microscopy (TCM), have been applied to U-10Zr fuel samples irradiated in EBR-II and FFTF to gain thorough understanding of fuel microstructures and property evolutions. Energy dispersive spectroscopy (EDS) and TEM images of fuel samples were used to analyze element composition and identify new phases. EDS images show the distributions of chemical elements, and TEM images demonstrate samples' surface microstructures. The EDS and selective area electron diffraction patterns (SAED) in a TEM have successfully identified different phases (crystal structures and compositions) within irradiated U-10Zr[8] in the nano-meter scale (see Figure 2).

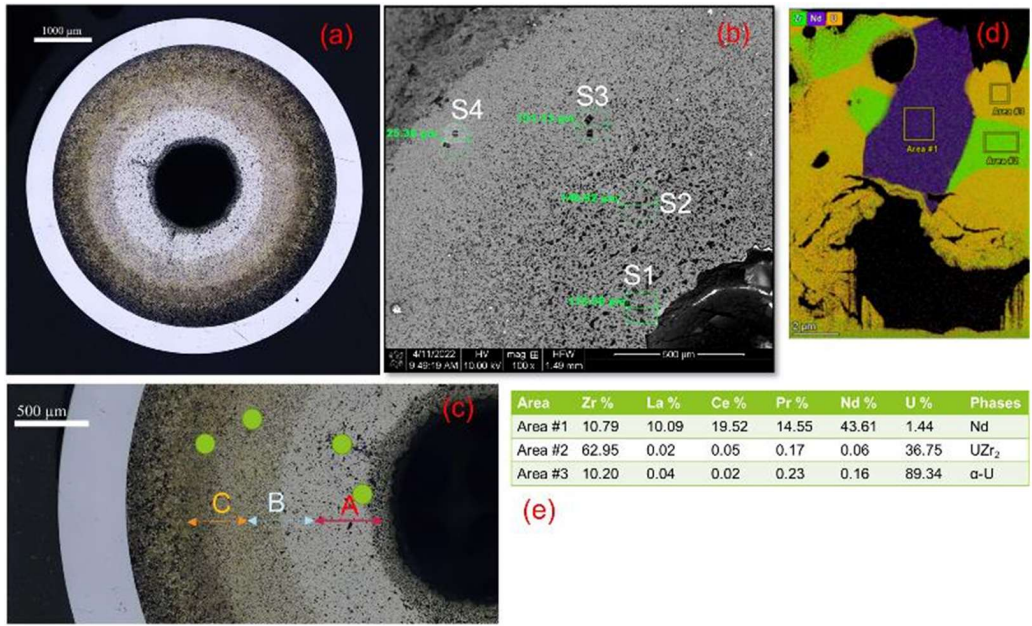

Figure 2. The micrographic cross section of a U-10Zr irradiated to 4.3% fission per initial metal atom (FIMA) burnup (a) and TEM compositions of elements study at different locations on the fuel (b-e). TEM-FIB samples of S1~S4 (d). Energy-Dispersive X-ray Spectroscopy (EDS) image of location S4 (d). The corresponding composition of elements (e). New microstructure $UZr_2$ is indicated as the area 2 in (d) and (e).

The purpose of this study was to further understand the Zr redistributions and microstructures' distributions of two advanced annular U-10Zr fuels. The irradiation condition of the two advanced annular U-10Zr fuels is shown in Table 1. The optical image of the AF2 is shown in Figure 2(a). Seven types of features exist on the fuel: 1) four phases, $\alpha$-Uranium matrix ($\alpha$-U), rich U with Zirconium (Zr) matrix phase, noted as (U, Zr) matrix, pure Zr, rich Zr with U, noted as $UZr_2$; and 2) three types of bubbles (isolated bubble, connected empty bubble, and connected bubble with fission products inside). As shown in Figure 2(c), the fuel cross-section could be separated into three regions with different compositions of elements. Three types of bubbles are major microstructural features in the (U, Zr) matrix area A. Three types of bubbles, (U, Zr) matrix and phase $UZr_2$, consist in the area B. The rest of the regions mainly contain pores and $UZr_2$ in the $\alpha$-U matrix. EDS results provided the elemental mapping of

U, Zr, and fission products. This study proposes a new framework to extract statistics of the seven features, using the SEM and EDS from six sample regions along the radial direction from the hot center region to the colder region close to the cladding region.

*Table 1. Information of the two advanced U-10Zr fuels*

| Fuel ID | Alloy | Fuel Form | Bond Material | Nominal Smear Density | Burnup | Temperature | FCCI |
|---|---|---|---|---|---|---|---|
| AF1 | U-10Zr | Annular | Helium | 55% | 3.3% | 540~600ºC | High |
| AF2 | U-10Zr | Annular | Helium | 55% | 4.3% | 600ºC | Low |

As shown in Figure 2, multiple phases were present even in the nano-meter scaled area. The EDS in scanning electron microscope (SEM) can provide elemental mapping in the micron-scaled area, but due to the complexity, it is impossible to derive the phase information from pure SEM/EDS. It is a challenge to bridge this phase information in a nano-meter scaled area from TEM, with the micron-scaled information from SEM/EDS, in order to obtain more meaningful statistical insights. In this study, EDS images were collected at six representative locations for a single fuel cross-section (see Figure 1). Additionally, TEM images from four locations were collected to identify the phases (see Figure 2 (b)-(d)). Moreover, a partial cross-section image is generated by stitching 587 high resolutions Back Scattered Electron (BSE) images. In Section 2.2, image processing techniques have been used to identify phases from EDS mapping with the aid from TEM results for the first time in U-10Zr.

Figure 3 shows two examples of SEM and EDS images, and there exists visual correlations between the two image modalities. The original EDS images used the RGB color model, and the color brightness has a linear relationship with the concentration of elements, (i.e., the brighter a pixel is, the higher concentration an element has). The first row of Figure 3 shows the 'Region 1' SEM image patch and its corresponding three EDS images (Zr, U and Nd). The second row is the images of 'Region 6.' By using EDS images, this study was able to generate partially annotated gas bubbles, identify new material phases, and calculate the statistics.

## 2.2. Phase Identification and Analysis

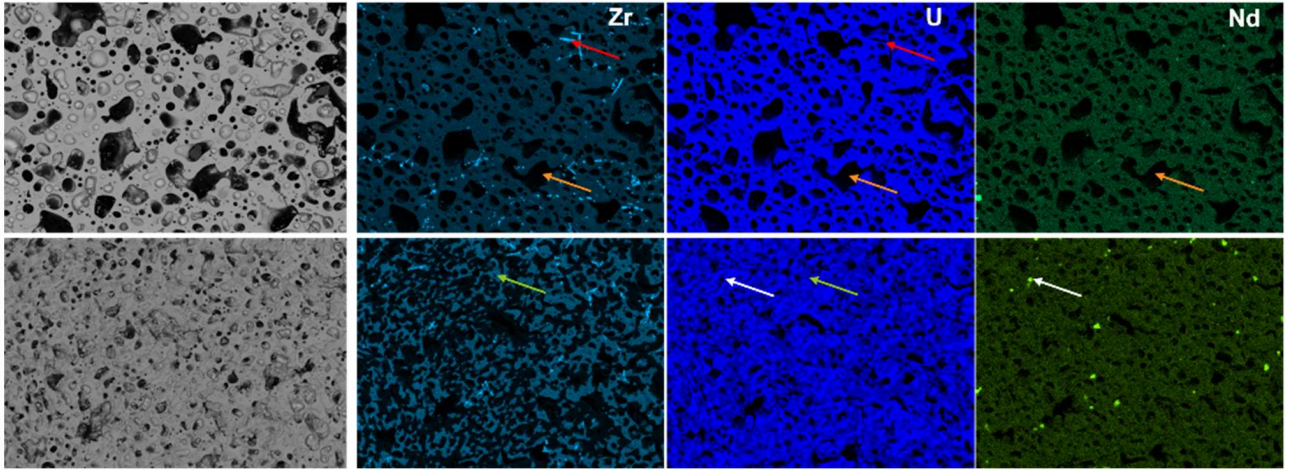

Figure 3. EDS images with elements composition information. Location with red arrow: pure zirconium (>90% Zr, <10% U); orange arrow: pores/bubbles; green arrow: $UZr_2$ (>50% Zr, <45% U); white arrow: Nd.

The pure Zr, $UZr_2$ and Nd phases can be easily obtained from EDS images (see the colored arrows in Figure 3). The pure Zr, pure Nd, (U, Zr) matrix, and α-U matrix are defined as the pixels with relative higher intensities in the corresponding images. The phase $UZr_2$ is defined as moderate intensities in both Zr and U images. The pores/bubbles appear relatively dark in SEM images. A multiple threshold segmentation method, which helps in separating the pixels into different groups according to their intensity level, is used to generate the initial annotated images of pure Zr, (U, Zr) matrix, $\alpha$-U matrix, $UZr_2$, Nd, and pores. $T_{k1} \sim T_{k3}$ (k can be Zr, U, or Nd) are the threshold values generated by the method on Zr, U and Nd EDS images. For an instance, in Zr EDS image, the pixels are separated into four groups based on the intensities as shown in Equation (1).

$$G_{Zr}(i,j) = \begin{cases} Pure\ Zr, & if\ I_{Zr}(i,j) \geq \max(T_{zr}, \delta_1 * 255) \\ No\ Zr\ element, & if\ I_{Zr}(i,j) < T_{zr} \\ UZr_2, & if\ I_{Zr}(i,j) > T_{zr3}\ and\ I_{Zr}(i,j) < \max(T_{zr1}, \delta_1 * 255) \\ (U,Zr)\ matrix\ or\ \alpha - U\ matrix, & others \end{cases} \quad \text{(Eqn. 1)}$$

$$G_U(i,j) = \begin{cases} High\ U, & if\ I_U(i,j) \geq \max(T_{u1}, \delta_2 * 255) \\ No\ U\ element, & if\ I_U(i,j) < T_{u2} \end{cases} \quad \text{(Eqn. 2)}$$

$$G_{Nd}(i,j) = \begin{cases} High\ Nd, & if\ I_{Nd}(i,j) \geq \max(T_{nd}, \delta_3 * 255) \\ No\ Nd\ element, & if\ I_{Nd}(i,j) < T_{nd2} \end{cases} \quad \text{(Eqn. 3)}$$

$$Pores = No\ Zr, U, or\ Nd\ in\ corresponding \quad \text{(Eqn. 4)}$$

where: $I_k$ is the corresponding EDS image; $G_k$ is the corresponding segmentation result; and k can be Zr, U, or Nd.

Pure Zr is assumed to be more than 90% Zr, alpha-U more than 80% Uranium, and high Nd more than

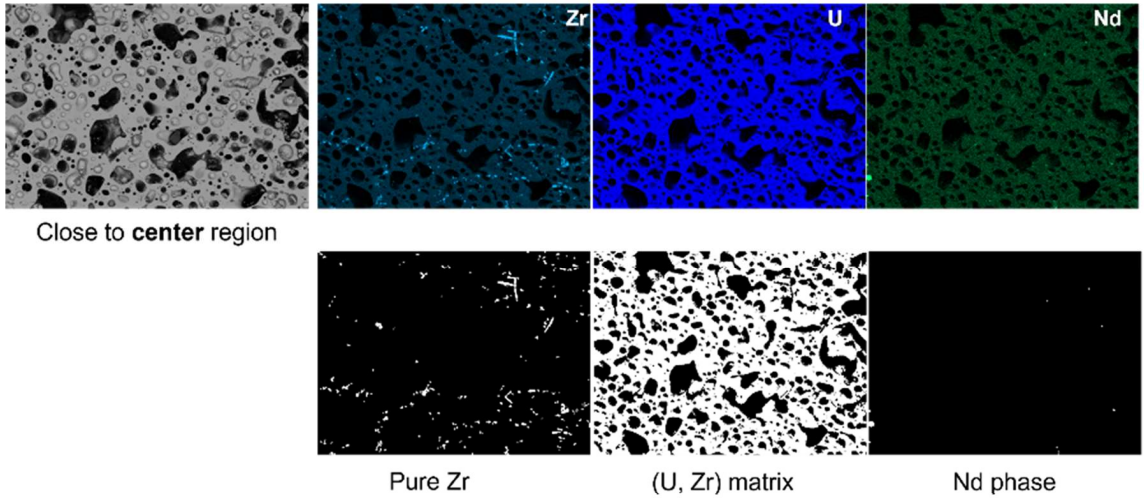

Figure 4. Region 1 (close to the hot fuel center) annotated Pure Zr, U, and Nd images.

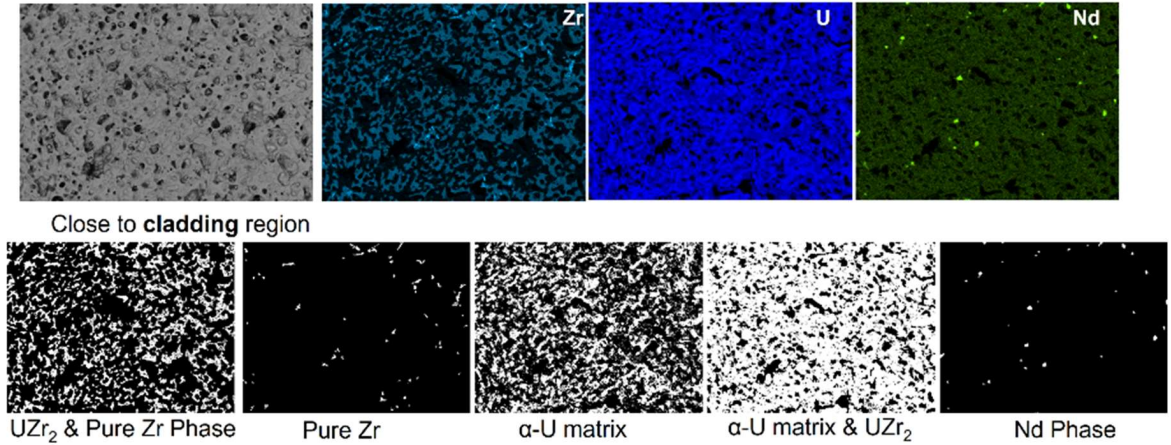

Figure 5. Region 6 (close to the cold rim) annotated Pure Zr, α-U matrix, UZr$_2$, and Nd images.

50% Nd in the composition. Therefore, we set $\delta_1$ as 0.9, $\delta_2$ as 0.8, and $\delta_3$ as 0.4 in Equations (1)-(4). The annotated images generated only based EDS images are shown in Figures 4 and 5 using two showcases of the Region 1 and Region 6. The distributions of six types of compositions based on analyzing the EDS images of six different locations on the fuel are shown in Table 2.

Table 2. Distributions of five types of compositions and porosities.

| Region* | α-U | Nd | Pure Zr | UZr$_2$ | (U, Zr) | Porosity |
|---|---|---|---|---|---|---|
| 1 | 0.000% | 0.036% | 1.291% | 0.000% | 71.659% | 27.041% |
| 2 | 0.000% | 0.342% | 1.225% | 0.000% | 69.483% | 29.184% |
| 3 | 0.000% | 0.317% | 1.182% | 0.000% | 76.727% | 22.021% |
| 4 | 0.000% | 0.646% | 1.123% | 0.000% | 76.532% | 22.202% |
| 5 | 55.399% | 0.000% | 0.704% | 35.860% | 0.000% | 8.036% |
| 6 | 57.590% | 0.122% | 0.072% | 36.512% | 0.000% | 5.710% |

* -The region is illustrated in Figure 7(a).

Three experts were involved in determining the annotated results with guidance from the TEM results. The multi-threshold method generated accurate results on the phases pure Zr, high Nd, and UZr$_2$. However, the challenging task was to annotate the pores accurately only based EDS images. Due to the height or angle factors when collecting the EDS images, some pores are not visible in the EDS images, as shown in Figures 6(a) and (b) marked with red circles.

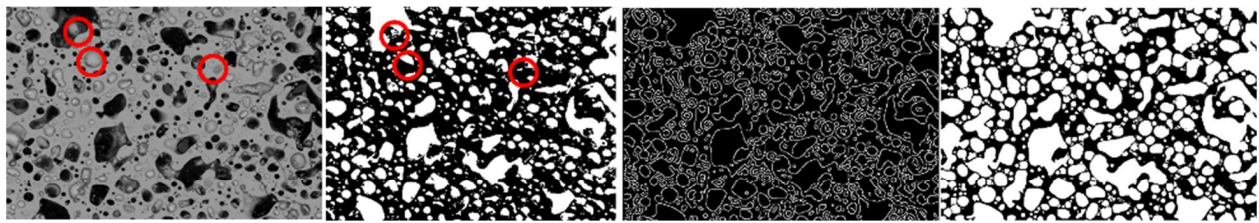

a) Original.   b) Dark regions in Uranium EDS.   c) Edge detection.   d) Annotated image.

Figure 6. Annotated image generation.

To verify and solve the problem, an edge detection method was employed to the original SEM image of the Region 1 first as shown in Figure 6(c), then the expert annotated the missing parts to generate the final annotated image as shown in Figure 6(d). The porosity of the annotated result of Region 1 is around 50.3%, which was distinct from approximately 27% in the 1st row, last column of Table 2. The verification results revealed EDS-SEM measurements inside the pores may be incorrect since these images reflected the lower surface element information under the pores. The annotated image will help demonstrate the distributions of microstructures along the cross-section from hot region to cladding regions. Additional results are shown in Figure 7. Figure 7(a) illustrates the six sample locations where EDS measurements were collected along the fuel cross-section. Sample regions 1-4 are located in area A of Figure 1(c) and Sample regions 5 and 6 are located in the area C of Figure 1(c). Figure 7(b) shows the relative areal percentage of four phases, pure Zr, $UZr_2$, (U, Zr) matrix, and α-U. The areal percentage is representative of the volumetric percentage if the whole fuel pin is irradiated under the same condition. The sample regions 1-4 contain <2% pure Zr and about 98% (U,Zr) matrix. The pure Zr content increases slightly towards the fuel center, from 1.4% to 1.8%, which matches previous observation that Zr migrates to the fuel center.[6] Sample regions 5 and 6 are consistent and contain about 39% $UZr_2$ and 61% α-U, which is very different from AF-1 in that the $UZr_2$ phase formed in the fuel center region instead.[8,18] The presence of $UZr_2$ created new interfaces that can stop the movement of the lanthanides,[18] so the appearance of $UZr_2$ close to the cladding may serve as a barrier for lanthanide transportation to reach cladding, another possible reason why AF1 has worse FCCI.

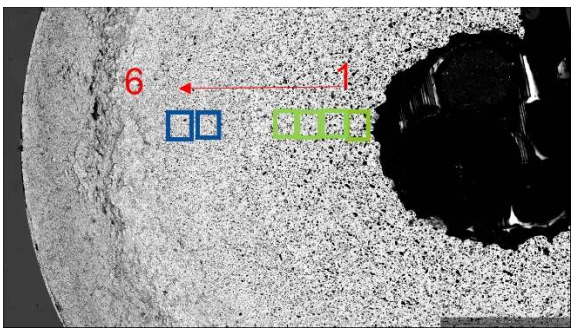

6 sample regions

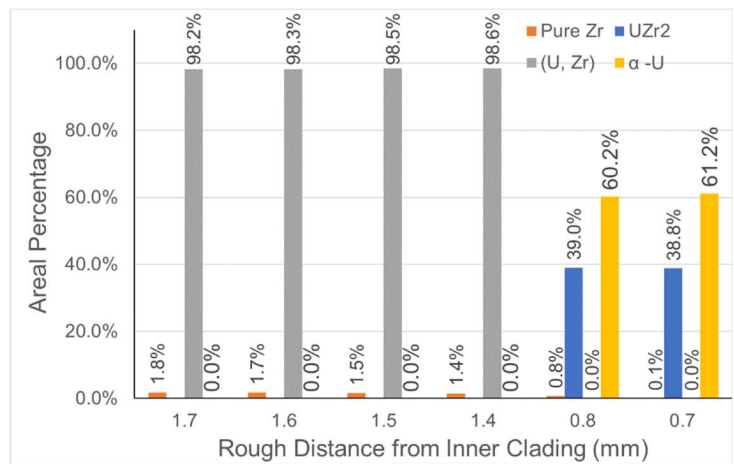

(a) (b)

Figure 7. (a) The illustration of the six locations where the EDS were collected. (b) The phase information at the six locations. Note: The y-axis of (b) is the areal percentage from the EDS and SEM. In the statistic result, the regions with other elements or pores were counted to the matrix, either (U, Zr) or α-U.

### 2.3. Bubble Detection and Classification

Xu, et.al proposed a workflow to detect the bubbles and then classify the bubbles into three categories: connected bubbles with fission products, connected empty bubbles, and isolated bubbles.[19] The method mainly contained four steps: detect the darker and brighter bubbles using two different thresholds; segment the bubbles' boundary using morphological operators; develop a decision tree model to classify the bubbles into three categories based on studying approximately 800 annotated bubbles; and divide the cross-section into 11 rings and generate the distributions based on bubbles' different properties.[18]

In this paper, a partial cross-section stitched image with image size 12619×18401 was generated, shown in Figure 8(a). A full fuel cross-section, with radius 48633 pixels and image size 97852×97266 pixels, was created by calculating the center and radius of the cross-section based on the cladding edge, see Figure 8(b). As shown in Figure 8(c), the fuel cross-section could be separated into three regions with different compositions of microstructures. Three types of bubbles are major microstructures in area A, which is the (U, Zr) matrix area. Three types of bubbles, α-U and $UZr_2$, are consist in area B. The rest of the regions mainly contains pores, UZr$_2$ and α-U. Obviously, the composition of UZr$_2$ is the key to identifying different areas. Exploration of the distributions of intensities under different regions, as show in Figure 9(a) and (b), found the peak with an intensity value greater than 200 is the essential measurement of UZr$_2$, shown as the red arrow in Figure 9(b). The more UZr$_2$, the higher the frequency value of the peak. Therefore, to compare the distributions in [17], we generated the area A with a radius range [665 μm, 1,065μm] from the fuel center and detected the bubbles using the method in [17]. The result is shown in Figure 9(c) with more than 17,700 bubbles.

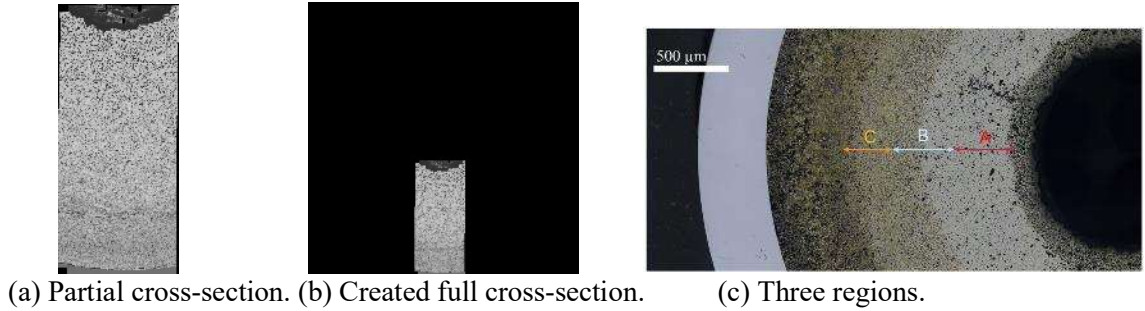

(a) Partial cross-section. (b) Created full cross-section. (c) Three regions.

Figure 8. Fuel Cross-section generation.

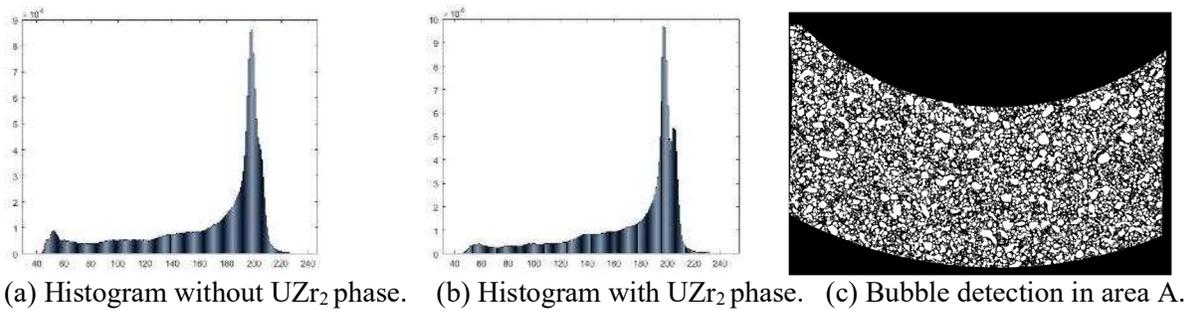

(a) Histogram without UZr$_2$ phase. (b) Histogram with UZr$_2$ phase. (c) Bubble detection in area A.

Figure 9. Background removal result.

A decision tree model was designed to classify the bubbles into three types by extracting 18 features, including the bubble size (F1), intensity histogram (F2 - F14), mean intensity (F15), the standard deviation of intensity (F16), intensity range (F17), and the shape convexity (F18) which is defined as the ratio of area over convex area.[18] The same model was applied to the bubble detection result of area A, shown in Figure 10.

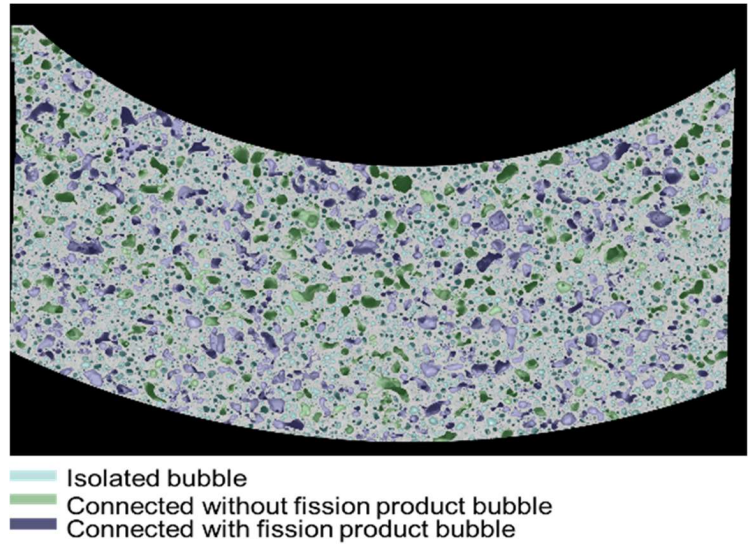

Figure 10. Bubble classification in the part of area A.

### 2.4. Bubble Statistics and its Implication on Nuclear Fuel Performance

The effect of fission gas bubbles on nuclear fuel performance has been extensively studied and focused on its impact on fuel swelling.[6] The gaseous fission product atoms have a limited solubility inside fuel matrix and tend to precipitate out in bubble form. Before releasing into the plenum or the void space inside the fuel rod, the fission gas bubbles grow with burnup as more gaseous fission products are produced by the fission reaction. When there is a preexisting bubble nearby, the growing bubble is expected to combine with the neighboring bubble to reduce the bubble surface area and to lower surface energy.

Accurate bubble statistics is the first step to derive the effect of gas bubbles on fuel performance. The bubble statistics of a total of approximately 50,000 bubbles over an area of 20 mm$^2$ for the AF1 U-10Zr fuel and approximately 17,000 bubbles over an area of 0.4 mm$^2$ for the AF2 U-10Zr fuel, are shown in Figures 11 and 12. Based on size, the bubbles are separated into large bubbles (area coverage >205 μm$^2$, corresponding to diameter >16.2 μm of round-shaped bubbles, in which $diameter = 2 \times \sqrt{area/\pi}$), the intermediate bubbles (32 < area ≤ 205 μm$^2$, or equivalent 6.4 < diameter ≤ 16.2 μm of round-shaped bubbles), and small bubbles (area ≤32 μm$^2$, or equivalent diameter ≤6.4 μm of round-shaped bubbles).

On the other hand, the bubbles are classified into three categories based on the bubble interconnection: isolated bubble, connected without fission products, and connected with fission products.

Figure11(a)(b) and (d)(e) are an overview of the fuel cross section from both metallography and SEM for AF1 and AF2 U-10Zr fuels, respectively. Figure 11(c) shows the porosity contribution of AF1 from the three bubble sizes (large, intermediate, and small) as a function of the distance from the inner cladding, with each column containing the porosity information from the eccentric annulus (ring) with each annular thickness about 0.08 mm. Figure 11(f) shows the porosity contribution of AF2 in the area A (U-Zr matrix) of Figure 8(c). The AF1 and AF2 fuels have very different microstructural features. In the high Zr region of the two fuels (fuel center), AF2 has higher porosity than AF1 (nearly 45% to 50% porosity in AF2, 25% to 40% porosity in AF1), however, large bubbles in AF1 cover more than 85% among the porosity, while AF2 is dominated by small and medium size bubbles, the number of the large bubbles is less than 20%. The quantitative results matched the observation that large pores located most of the fuel cross-section of AF1 and not many obvious large pores were observed in AF2 as shown in Figure 11(a) and (d). Moreover, the small gas bubbles in AF2 created overall higher porosity but seems to cause less fuel swelling, while the interconnected gas bubbles may be responsible for more severe fuel swelling and more filling of the central holes in AF1.

In addition, the AF2 has more isolated and connected without lanthanides pores as shown in Figure 12(b), while the AF1 has much more connected pores with lanthanides inside as shown in Figure 12(a). These statistical data matches well with the observations that AF2 has much less FCCI, because connected pores with lanthanides are the pathway for lanthanides transported to the cladding. As expected, lanthanides move down the temperature gradient along the interconnected bubbles (channel) to the inner cladding surface. The small or isolated pores contribute minimal to the lanthanide movement, therefore, there are much fewer pores with lanthanides for AF2 as well as fewer lanthanides transported through the interconnected bubbles (shortcut) to the inner cladding for FCCI.

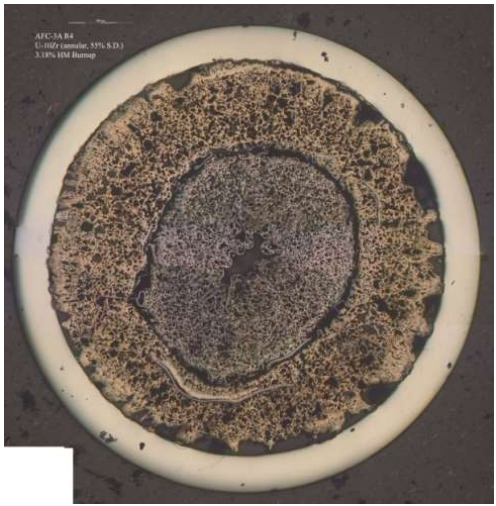

(a)

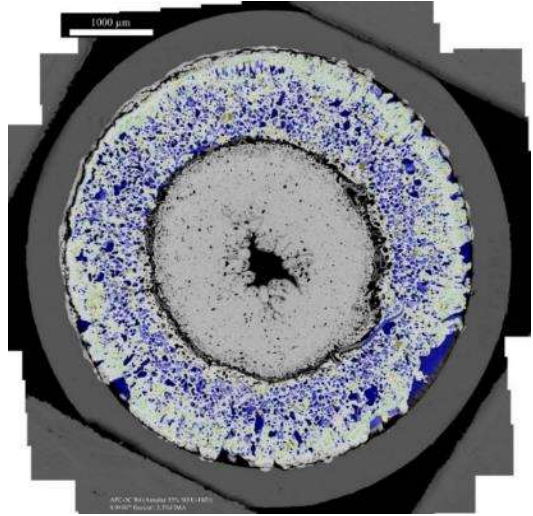

(b)

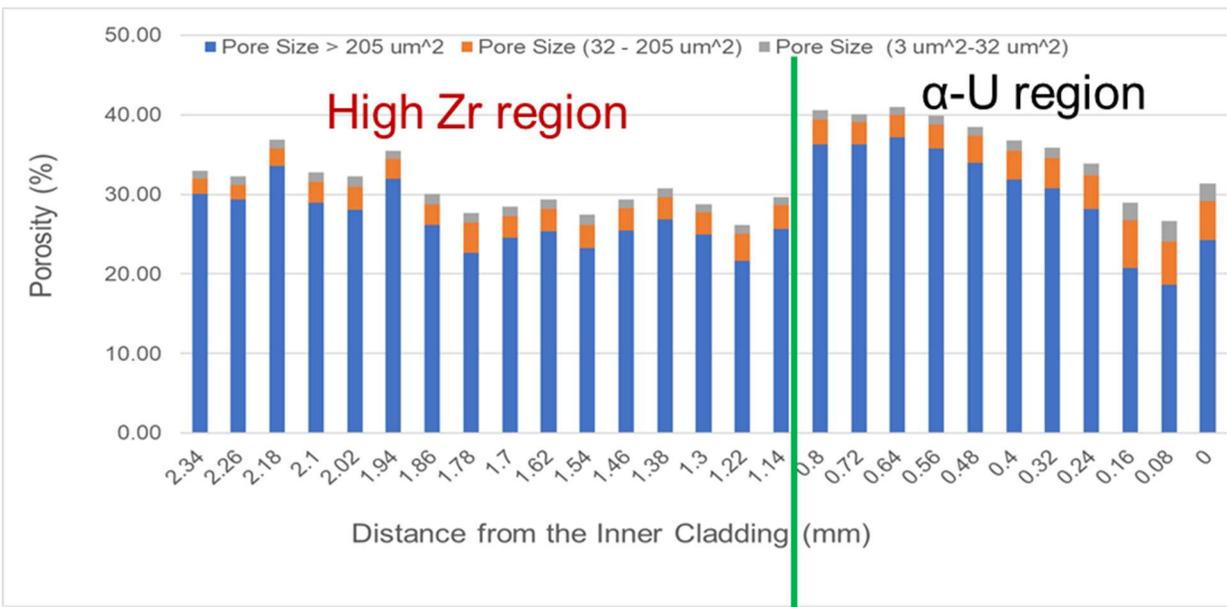

(c)

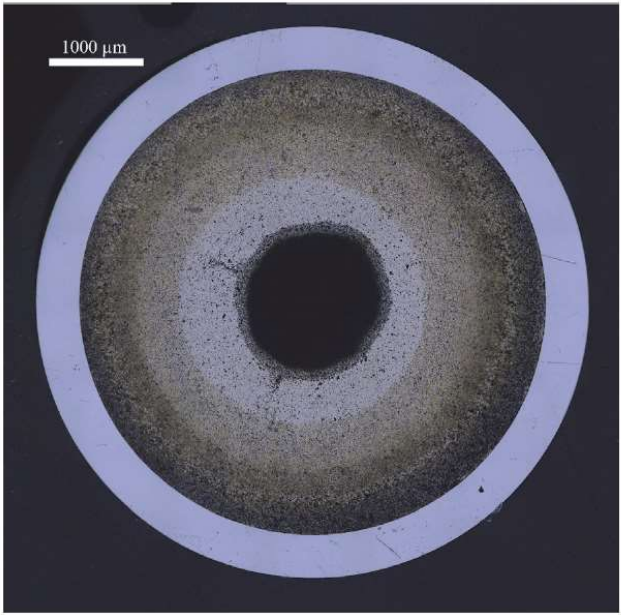
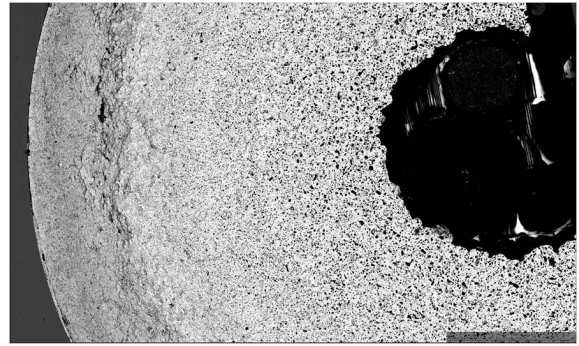

(d) (e)

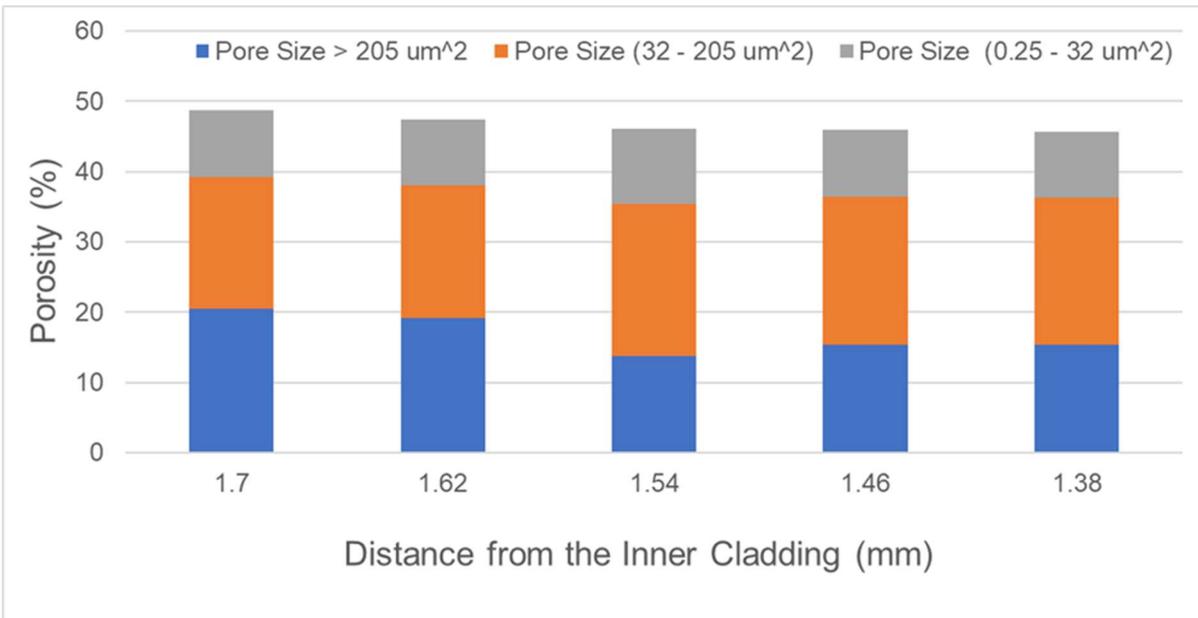

(f)

Figure 11. (a) Metallography and (b) SEM cross-section of U-10Zr (AF1). (c) The porosity distribution as function of distance from the inner cladding (AF1). (d) Metallography and (e) SEM cross-section of U-10Zr (AF2). (f) The porosity distribution as function of distance from the inner cladding (AF2) in the high zirconium region.

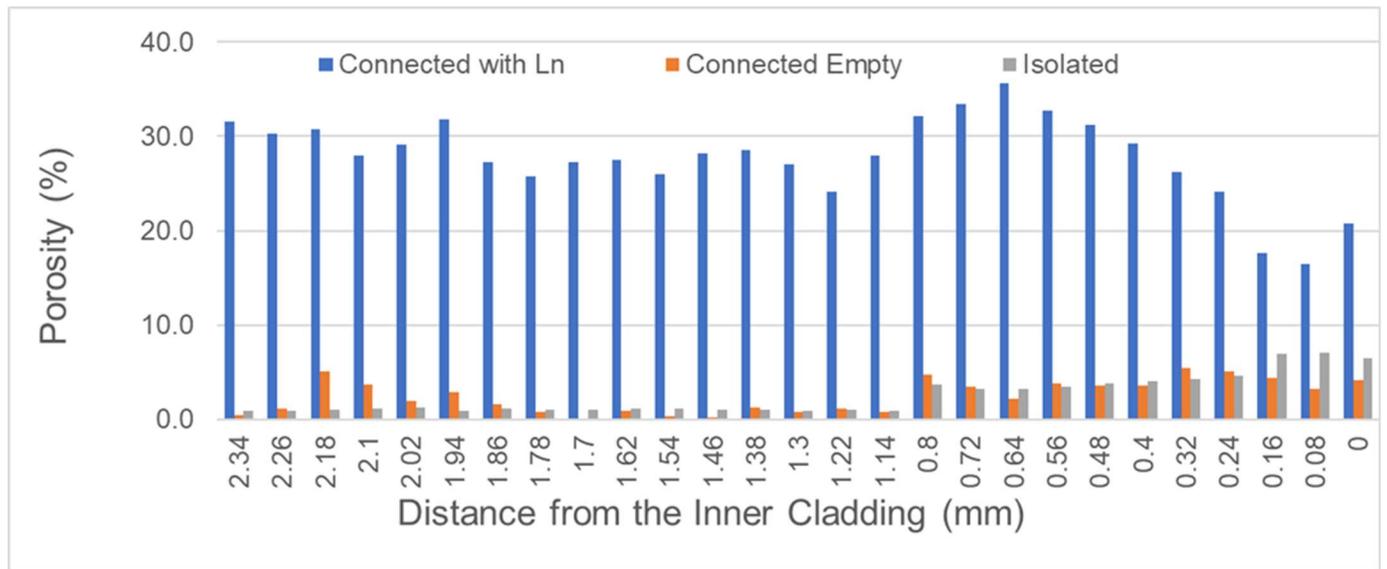

(a)

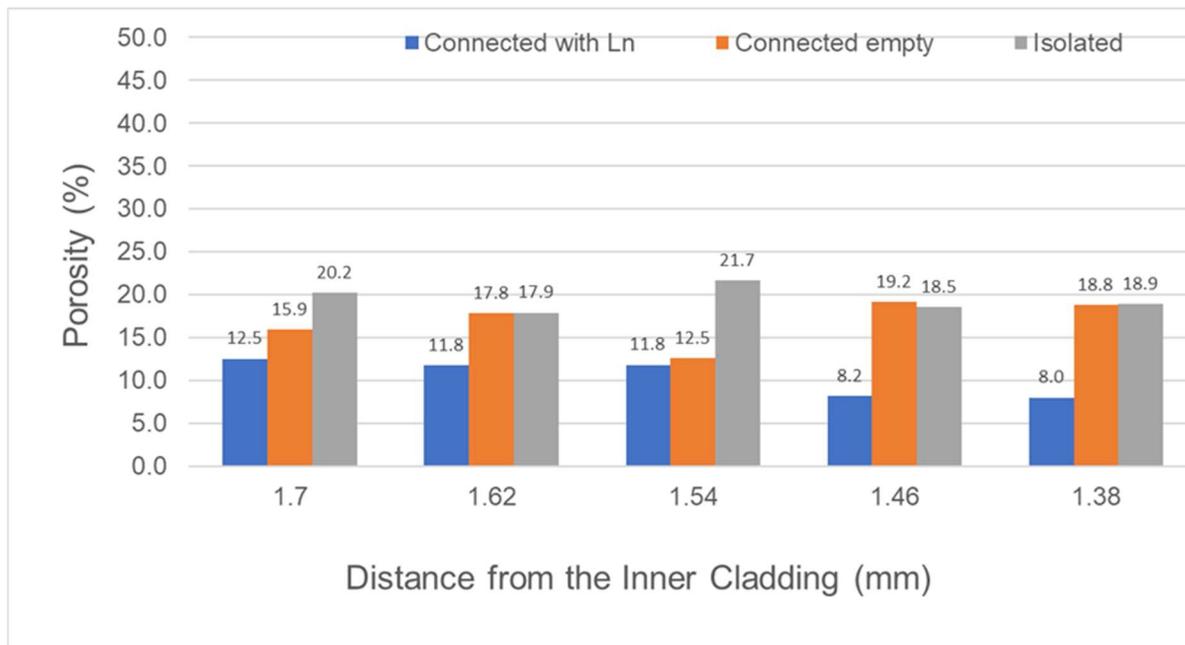

(b)
Figure 12. The porosity distribution of three bubble (pore) types as a function of distance from the inner cladding.

## 3. Conclusion

This work developed a new framework with integrating multi-scaled images (nano-meter to micrometer scales) and multiple sources (SEM, EDS, FIB, and TEM) to achieve the mechanistic understanding of fuel performance (i.e., identifying the phase $UZr_2$ existing, and revealing Zr redistribution along thermal gradient from center region to cladding, and the porosity and other phases' distributions). Moreover, this work demonstrated the applicability of the ML model that is built from AF1 to classify the fission gas bubbles' categories by successfully applying the model to the new advanced U-10Zr fuel, AF2.[17] Additionally, the following conclusive findings were obtained:

1. The $UZr_2$ phase was observed in the AF2 and AF1 advanced U-10Zr fuels. $UZr_2$ locates at the regions close to the cladding of AF2, while $UZr_2$ exists in the region close to center in AF1.
2. In the High Zr region of the two fuels, AF2 has a higher porosity than AF1 (nearly 45% to 50% porosity in AF2, and 25% to 40% porosity in AF1).
3. Large bubbles in AF1 cover more than 85% among the porosity; while AF2 is dominated by small and medium size bubbles, the number of the large bubbles is less than 20%.
4. AF1 has twice the among of larger and interconnected bubbles than AF2. This finding agrees with experimental results showing stronger FCCI and small center holes for filling in AF1 than AF2.
5. The model linking the information from SEM, EDS, STEM, and TEM was proposed for the first time to detect pores, phases, compositions, and the corresponding distributions.
6. Generating the distributions of content of Zr, porosity, and other phases along the thermal gradient from center region to cladding of AF2 is beyond human's capability. The new framework accelerates the quantitative analysis and will eventually accelerate quantitative research of fuel qualification.

The Machine Learning model provides quantitative data, which is extremely difficult to derive manually, to investigate the performance of advanced fuels including fuel swelling, lanthanide migration, and FCCI. With this data we are building the bridge between the experimental data and the mechanistic understanding of fuel behavior to accelerate fuel qualification.

## 4. Acknowledgements

This work was supported by the U.S. Department of Energy, Office of Nuclear Energy under DOE Idaho Operations Office Contract DE-AC07-05ID14517 and LDRD project of 22A1059-094FP. The authors also acknowledge the support of DOE Advanced Fuel Campaign on the experimental data collection. Accordingly, the U.S. Government retains and the publisher, by accepting the article for publication, acknowledges that the U.S. Government retains a nonexclusive, paid-up, irrevocable, worldwide license to publish or reproduce the published form of this manuscript or allow others to do so, for U.S. Government purposes. Insightful discussion with Larry K Aagesen Jr. is also greatly appreciated.

## 5. Disclaimer

## 6. Data Availability Statement

The data that support the findings of this study are available from the corresponding author upon reasonable request.

## 7. References

1  Schmidt, J., Marques, M. R. G., Botti, S. & Marques, M. A. L. Recent advances and applications of machine learning in solid-state materials science. *npj Computational Materials* **5**, 83, doi:10.1038/s41524-019-0221-0 (2019).
2  Wang, A. Y.-T., Murdock, R. J., Kauwe, S. K., Oliynyk, A. O., Gurlo, A., Brgoch, J., Persson, K. A. & Sparks, T. D. Machine Learning for Materials Scientists: An Introductory Guide toward Best Practices. *Chemistry of Materials* **32**, 4954-4965, doi:10.1021/acs.chemmater.0c01907 (2020).
3  Morgan, D., Pilania, G., Couet, A., Uberuaga, B. P., Sun, C. & Li, J. Machine learning in nuclear materials research. *Current Opinion in Solid State and Materials Science* **26**, 100975, doi:https://doi.org/10.1016/j.cossms.2021.100975 (2022).
4  Allen, T., Busby, J., Meyer, M. & Petti, D. Materials challenges for nuclear systems. *Materials Today* **13**, 14-23, doi:https://doi.org/10.1016/S1369-7021(10)70220-0 (2010).
5  Janney, D. E. & Hayes, S. L. Experimentally Known Properties of U-10Zr Alloys: A Critical Review. *Nucl Technol* **203**, 109-128, doi:10.1080/00295450.2018.1435137 (2018).
6  Carmack, W. J., Porter, D. L., Chang, Y. I., Hayes, S. L., Meyer, M. K., Burkes, D. E., Lee, C. B., Mizuno, T., Delage, F. & Somers, J. Metallic fuels for advanced reactors. *J Nucl Mater* **392**, 139-150, doi:10.1016/j.jnucmat.2009.03.007 (2009).

7       Ogata, T. in *Comprehensive Nuclear Materials (Second Edition)*   (eds Rudy J. M. Konings & Roger E. Stoller)  1-42 (Elsevier, 2020).
8       Yao, T. K., Capriotti, L., Harp, J. M., Liu, X., Wang, Y. C., Teng, F., Murray, D. J., Winston, A. J., Gan, J., Benson, M. T. & He, L. F. alpha-U and omega-UZr2 in neutron irradiated U-10Zr annular metallic fuel. *J Nucl Mater* **542**, doi:ARTN 152536
10.1016/j.jnucmat.2020.152536 (2020).
9       Benson, M. T., Harp, J. M., Xie, Y., Yao, T. K., Tolman, K. R., Wright, K. E., King, J. A., Hawari, A. I. & Cai, Q. S. Out-of-pile and postirradiated examination of lanthanide and lanthanide-palladium interactions for metallic fuel. *J Nucl Mater* **544**, doi:ARTN 152727
10.1016/j.jnucmat.2020.152727 (2021).
10      Salvato, D., Liu, X., Murray, D. J., Paaren, K. M., Xu, F., Pavlov, T., Benson, M. T., Capriotti, L. & Yao, T. Transmission electron microscopy study of a high burnup U-10Zr metallic fuel. *J Nucl Mater* **570**, 153963, doi:https://doi.org/10.1016/j.jnucmat.2022.153963 (2022).
11      Aitkaliyeva, A. Recent trends in metallic fast reactor fuels research. *J Nucl Mater* **558**, 153377, doi:https://doi.org/10.1016/j.jnucmat.2021.153377 (2022).
12      Matthews, C., Unal, C., Galloway, J., Keiser, D. D. & Hayes, S. L. Fuel-Cladding Chemical Interaction in U-Pu-Zr Metallic Fuels: A Critical Review. *Nucl Technol* **198**, 231-259, doi:10.1080/00295450.2017.1323535 (2017).
13      Keiser, D. D. Fuel cladding chemical interaction in metallic sodium fast reactor fuels: A historical perspective. *J Nucl Mater* **514**, 393-398, doi:10.1016/j.jnucmat.2018.09.045 (2019).
14      Zhang, J. & Taylor, C. Studies of Lanthanide Transport in Metallic Fuel. Report No. 14-6482, (The Ohio State University, 2018).
15      Bauer, T. H. & Holland, J. W. In-Pile Measurement of the Thermal-Conductivity of Irradiated Metallic Fuel. *Nucl Technol* **110**, 407-421, doi:Doi 10.13182/Nse110-407 (1995).
16      Yun, D., Yacout, A. M., Stan, M., Bauer, T. H. & Wright, A. E. Simulation of the impact of 3-D porosity distribution in metallic U-10Zr fuels. *J Nucl Mater* **448**, 129-138, doi:10.1016/j.jnucmat.2014.02.002 (2014).
17      Cai, L., Xu, F., Di Lemma, F. G., Giglio, J. J., Benson, M. T., Murray, D. J., Adkins, C. A., Kane, J. J., Xian, M., Capriotti, L. & Yao, T. Understanding fission gas bubble distribution, lanthanide transportation, and thermal conductivity degradation in neutron-irradiated α-U using machine learning. *Materials Characterization* **184**, 111657, doi:https://doi.org/10.1016/j.matchar.2021.111657 (2022).
18      Liu, X., Capriotti, L., Yao, T., Harp, J. M., Benson, M. T., Wang, Y., Teng, F. & He, L. Fuel-cladding chemical interaction of a prototype annular U-10Zr fuel with Fe-12Cr ferritic/martensitic HT-9 cladding. *J Nucl Mater* **544**, 152588, doi:https://doi.org/10.1016/j.jnucmat.2020.152588 (2021).
19      Xu, F., Cai, L., Salvato, D., Dilemma, F., Giglio, J. J., Benson, M., Murray, D. J., Adkins, C. A., Kane, J. J., Xian, M., Capriotti, L. & Yao, T. Understanding Fission Gas Bubble Distribution and Zirconium Redistribution in Neutron-irradiated U-Zr Metallic Fuel Using Machine Learning. *Microscopy and Microanalysis* **28**, 82-83, doi:10.1017/S1431927622001234 (2022).